\documentclass[12pt]{article}

\usepackage{graphicx}
\usepackage{amsmath,amssymb}
\usepackage{bm}
\usepackage{graphicx, color}
\usepackage{wrapfig}
\usepackage{url}
\usepackage{comment}
\begin{document}
\title{
\begin{flushright}
\ \\*[-80pt] 
\begin{minipage}{0.2\linewidth}
\normalsize
KIAS-P17059\\
KIAS-Q17016\\
HUPD1706 
\\*[50pt]
\end{minipage}
\end{flushright}
{\Large \bf 
Light flavon signals at electron - photon colliders
\\*[20pt]}}

\author{ 
\centerline{
~Yu~Muramatsu$^{1}$\footnote{E-mail address: yumura@mail.ccnu.edu.cn},
~Takaaki~Nomura$^{2}$\footnote{E-mail address: nomura@kias.re.kr},
~Yusuke~Shimizu$^{3}$\footnote{E-mail address: yu-shimizu@hiroshima-u.ac.jp},~and
~Hiroshi~Yokoya$^{4}$\footnote{E-mail address: hyokoya@kias.re.kr}
 }
\\*[20pt]
\centerline{
\begin{minipage}{\linewidth}
\begin{center}
$^1${\it \normalsize 
Institute~of~Particle~Physics~and~Key~Laboratory~of~Quark~and~Lepton~
Physics~(MOE), Central~China~Normal~University,~Wuhan,~Hubei~430079,
~People's~Republic~of~China}\\
$^2${\it \normalsize 
School~of~Physics,~KIAS,~Seoul~130-722,~Republic~of~Korea} \\
$^3${\it \normalsize 
Graduate~School~of~Science,~Hiroshima~University,~Higashi-Hiroshima,~739-8526,~Japan}\\
$^4${\it \normalsize 
Quantum~Universe~Center,~KIAS,~Seoul~130-722,~Republic~of~Korea}\\
\end{center}
\end{minipage}}
\\*[50pt]}

\date{
\centerline{\small \bf Abstract}
\begin{minipage}{0.9\linewidth}
\medskip 
\medskip 
\small
Flavor symmetries are useful to realize fermion flavor structures 
in the standard model.
In particular, discrete $A_4$ symmetry is used to realize 
lepton flavor structures, and some scalars which are called flavon 
are introduced to break this symmetry.
In many models, flavons are assumed to be much heavier than 
the electroweak scale.
However, our previous work showed that flavon mass around 
100 GeV is allowed by experimental constraints in 
the $A_4$ symmetric model with residual $Z_3$ symmetry.
In this paper, we discuss collider search of such a light flavon $\varphi_T$.
We find that an electron - photon collision, as a considerable option at the  international linear collider, 
has advantages to search for the signals.
At the electron - photon collider flavons are 
produced as $e^-\gamma \to l^- \varphi_T$ and decay into 
two charged leptons.
Then we analyze signals of flavor-conserving final-state 
$\tau^+ \tau^- e^-$, and flavor-violating final-states 
$\tau^+ \mu^- \mu^-$ and $\mu^+ \tau^- \tau^-$ by carrying out 
numerical simulation.
For the former final-state, SM background can be strongly suppressed by 
imposing cuts on the invariant masses of final-state leptons.
For the later final-states, SM background is extremely small, 
because in the SM there are no such flavor-violating final-states.
We then find that sufficient discovery significance can be obtained,  
even if flavons are heavier than the lower limits from flavor physics.
\end{minipage}
}

\begin{titlepage}
\maketitle
\thispagestyle{empty}
\end{titlepage}

\section{Introduction}
The standard model (SM) particles are completed by discovery of the Higgs boson. 
However, the origin of the generation and flavor structure of the SM fermions is not clear. 
In order to explain the flavor structure of the SM fermions, we introduce flavor symmetries and scalar fields, so-called ``flavons".
After flavons take vacuum expectation values (VEVs), flavor symmetries are broken and SM fermions obtain flavor structure.

In the SM neutrinos are massless.
However neutrino oscillation experiments reveal neutrino mass squared differences and large lepton mixing angles~\cite{An:2012eh}-\cite{Abe:2013hdq}.
In order to explain large lepton mixing angles, many authors have been studying the lepton flavor structure 
by using non-Abelian discrete symmetries as flavor symmetry (See for review~\cite{Ishimori:2010au}-\cite{King:2014nza}.). 
The non-Abelian discrete flavor symmetry can easily derive the large lepton mixing angles, e.g. tri-bimaximal mixing (TBM) which is 
a simple paradigm of the lepton mixing matrix.
Indeed, Altarelli and Feruglio (AF) proposed an $A_4$ flavor model~\cite{Altarelli:2005yp,Altarelli:2005yx} 
which introduces gauge singlet flavons in addition to the $SU(2)$ doublet SM Higgs field. 
In the AF model, the lepton mixing matrix is the exact TBM one. 
However the observation of the non-zero reactor angle 
forces us to study the deviation from the TBM 
or study other flavor paradigms. In Ref.~\cite{Shimizu:2011xg}, the authors predicted the non-zero reactor angle with breaking TBM by adding an extra flavon to the AF model. 

In the experimental point of view, flavor symmetries have not been confirmed yet. 
Many authors have tried to predict the Dirac CP violating phase, Majorana phases, and effective mass for the neutrinoless double beta decay, which can be an indirect evidence for flavons. 
In Refs.~\cite{Holthausen:2012wz}-\cite{Varzielas:2015sno}, the authors discussed the mass restriction on the flavons which 
mix with the SM Higgs, from the lepton flavor violation (LFV) and collider physics. 
On the other hand in our previous work~\cite{Muramatsu:2016bda}, we studied experimental constraints for flavons which do not mix with the SM Higgs.
From LFV constraints the lower limit of the flavon masses is around 65 GeV.
Such a light flavon mass limit comes from the residual $Z_3$ symmetry.

Because of the light flavon mass limit we can expect direct flavon signals at colliders.
First we examine this possibility at the large hadron collider (LHC).
However as we will show it is hard to find flavon signals at the LHC.
Then we examine other possibilities at lepton colliders.
In particular we find that an electron - photon collider has many advantages in searching for flavon signals.
The photon beam is obtained from backscattered Compton photons~\cite{Ginzburg:1982yr, Ginzburg:1999wz}, and therefore the electron - photon collision could be realized at future lepton colliders.
Possibilities of the electron - photon collider have been discussed from 
a long time ago \cite{Ginzburg:1982yr, Ginzburg:1981, Ginzburg:1983}.
At the international linear collider (ILC), phenomenology in the electron - photon collision has been discussed \cite{Kanemura:2009, Yue:2010}.
In this paper we show that sufficient discovery significance can be obtained in the electron - photon collision at the ILC.

This paper is organized as follows:
In Section \ref{sec:2}, we summarize the modified AF model.
In Section \ref{sec:3}, we examine flavon signals at the LHC.
In Section \ref{sec:4}, we show advantages in searching for flavon signals in electron - photon collision at the ILC.
Moreover we show flavon signals in flavor conserving processes and flavor violating processes.
Summary and discussion are given in Section \ref{sec:7}. 

\section{$A_4$ flavor model}\label{sec:2}
In this section, we briefly summarize the modified AF model~\cite{Muramatsu:2016bda}. 
First of all, we discuss the mass of flavon $\phi _T$ which is triplet under the $A_4$ group and couples to the charged leptons.  
One of the solutions of the potential minimum in Ref.~\cite{Muramatsu:2016bda,Morozumi:2017rrg} leads that $\phi _T$ take the VEV $v _T$ as follows:
\begin{equation}
\langle \phi _T\rangle =v_T(1,0,0),\qquad v_T=\frac{3M}{2g},
\label{eq:alignment-vT}
\end{equation}
where $M$ is a mass parameter and 
$g$ is a trilinear coupling in the flavon potential.
By using the VEV in Eq.~(\ref{eq:alignment-vT}), 
we calculate the mass of the flavon $\phi _T$. 
We expand the flavon field around the VEV $v_T$ as 
\begin{equation}
\phi _T=\left (\phi _{T1},\phi _{T2},\phi _{T3}\right )\rightarrow \left (v_T+\varphi _{T1},\varphi _{T2},\varphi _{T3}\right ),
\label{eq:expanding-around-VEV}
\end{equation}
where $\varphi _{Ti}$ are complex scalar fields. 
Then, masses of the scalar fields $m_{\varphi _{Ti}}$ are obtained as 
\begin{equation}
(m_{\varphi _{T1}}^2,m_{\varphi _{T2}}^2,m_{\varphi _{T3}}^2)=(2M^2,8M^2,8M^2),
\end{equation}
and $\varphi _{Ti}$ do not mix each other.
Hereafter we assume that $v_T = 2m_{\varphi_{T1}} = m_{\varphi_{T2}} = m_{\varphi_{T3}}$ for simplicity.

Next we discuss the charged lepton sector.
The Lagrangian including SM Yukawa and flavon Yukawa interactions are written as follows~\cite{Muramatsu:2016bda}: 
\begin{equation}
\mathcal{L}_\ell =
y_e\left (\phi _T\bar l\right )e_Rh_d/\Lambda +
y_\mu \left (\phi _T\bar l\right )\mu _R h_d/\Lambda +
y_\tau \left (\phi _T\bar l\right )\tau _Rh_d/\Lambda +h.c.,
\end{equation} 
where $y_\alpha ~(\alpha =e,\mu ,\tau )$ are Yukawa couplings, $\Lambda $ is an $A_4$ cut-off scale, $h_d$ is $SU(2)$ doublet Higgs. 
The left-handed lepton doublets $l=(l_e, l_\mu, l_\tau)$ are assigned to triplet under the $A_4$ group, while the right-handed charged leptons $e_R$, $\mu _R$, and $\tau _R$ are assigned to singlet denoted as $\bf 1$, $\bf 1''$, and $\bf 1'$, respectively. 
After expanding $\phi _T$ around $v_T$ as in Eq.~(\ref{eq:expanding-around-VEV}) and 
taking the VEV of $SU(2)$ doublet Higgs $h_d$ denoted as $v_d$, 
the charged lepton mass term $\mathcal{L}_{\ell }^{\text{mass}}$ is written as
\begin{align}
\mathcal{L}_{\ell }^{\text{mass}}&=
\begin{pmatrix}
\bar e_L & \bar \mu _L& \bar \tau _L
\end{pmatrix}
\begin{pmatrix}
\frac{y_e v_d}{\Lambda }v_T & 0 & 0 \\
0 & \frac{y_\mu v_d}{\Lambda }v_T & 0 \\
0 & 0 & \frac{y_\tau v_d}{\Lambda }v_T
\end{pmatrix}
\begin{pmatrix}
e_R \\
\mu _R \\
\tau _R
\end{pmatrix}+h.c. \nonumber \\
&\equiv 
\begin{pmatrix}
\bar e_L & \bar \mu _L& \bar \tau _L
\end{pmatrix}
\begin{pmatrix}
m_e & 0 & 0 \\
0 & m_\mu & 0 \\
0 & 0 & m_\tau 
\end{pmatrix}
\begin{pmatrix}
e_R \\
\mu _R \\
\tau _R
\end{pmatrix}+h.c..
\end{align}
In our model, charged leptons in the interaction basis are equal to those in the mass basis. 
Therefore, there is no mixing in the charged lepton sector in the leading order level.
Moreover the charged lepton and flavon interaction term $\mathcal{L}_{\ell }^{\text{FY}}$ is obtained as 
\begin{align}
\mathcal{L}_{\ell }^{\text{FY}}&=
\begin{pmatrix}
\bar e_L & \bar \mu _L& \bar \tau _L
\end{pmatrix}
\begin{pmatrix}
\frac{m_e}{v_T} & 0 & 0 \\
0 & \frac{m_\mu}{v_T} & 0 \\
0 & 0 & \frac{m_\tau}{v_T}
\end{pmatrix}
\begin{pmatrix}
e_R \\
\mu _R \\
\tau _R
\end{pmatrix}
\varphi _{T1} \nonumber \\
&+
\begin{pmatrix}
\bar e_L & \bar \mu _L& \bar \tau _L
\end{pmatrix}
\begin{pmatrix}
0 & \frac{m_\mu }{v_T} & 0 \\
0 & 0 & \frac{m_\tau }{v_T} \\
\frac{m_e}{v_T} & 0 & 0
\end{pmatrix}
\begin{pmatrix}
e_R \\
\mu _R \\
\tau _R
\end{pmatrix}
\varphi _{T2} \nonumber \\
&+
\begin{pmatrix}
\bar e_L & \bar \mu _L& \bar \tau _L
\end{pmatrix}
\begin{pmatrix}
0 & 0 & \frac{m_\tau }{v_T} \\
\frac{m_e}{v_T} & 0 & 0 \\
0 & \frac{m_\mu }{v_T} & 0
\end{pmatrix}
\begin{pmatrix}
e_R \\
\mu _R \\
\tau _R
\end{pmatrix}
\varphi _{T3}+h.c..
\end{align}
We find that $\varphi_{T1}$ exchange does not induce flavor violation, 
while the other flavon exchanges induce flavor violation. 
We also find that the couplings are fixed by charged lepton masses 
except for $v_T$.

Before closing this section, we mention the lower bound of the flavon mass, which has been discussed in Ref.~\cite{Muramatsu:2016bda}.
After taking VEV of the flavon, the $A_4$ symmetry is broken down to the residual $Z_3$ symmetry. 
Thanks to the residual $Z_3$ symmetry, many lepton flavor violating decay modes are forbidden\footnote{
In some models with the $A_4$ symmetry, LFVs which come from $Z_3$-breaking effects are discussed \cite{Pascoli:2016wlt}.
}. 
Then, the mass lower bound of the flavon is found to be $60$~GeV which comes from the $\tau$ three-body decay mode $\tau^\pm \rightarrow \mu^\pm \mu^\pm e^\mp$.
In addition if we assume Yukawa coupling $y_\tau$ to be $\mathcal{O}(1)$ the cut off scale is typically $\mathcal{O}(10)$ TeV.

\section{Flavon signals at proton - proton collider}\label{sec:3}
In our previous paper \cite{Muramatsu:2016bda}, it is shown that in our model 
the flavon can be light, and therefore we expect flavon signals at colliders.
Hence we show flavon production cross section at the LHC.
In Table \ref{tab:FPCS} we summarize the flavon production cross section calculated in Ref. 
\cite{Muramatsu:2016bda}.
\begin{table}[ht]
\begin{center}
\begin{tabular}{c|cccccc} \hline \hline
Final-state & $\varphi_{T1}\tau^- \bar{\nu}_\tau$ & $\varphi_{T1}\tau^- \tau^+$ & 
$\varphi_{T2}\tau^- \bar{\nu}_\mu$ & $\varphi_{T2}\tau^- \mu^+$ & 
$\varphi_{T3}\tau^- \bar{\nu}_e$ & $\varphi_{T3}\tau^- e^+$ \\
Cross section [fb] & 2.2 & $1.5\times10^{-1}$ & $1.7\times10^{-5}$ & 
$8.4\times10^{-6}$ & $1.7\times10^{-5}$ & $8.4\times10^{-6}$ \\ \hline \hline 
\end{tabular}
\caption{
The production cross sections at the LHC for each flavon where 
$\sqrt{s}=14$ TeV and the flavon masses are 
$v_T = 2m_{\varphi_{T1}} = m_{\varphi_{T2}} = m_{\varphi_{T3}} = 65$ GeV.
In this calculation we use MadGraph with default momentum and rapidity cuts.
}
\label{tab:FPCS}
\end{center}
\end{table}
In this calculation we set $\sqrt{s}=14$ TeV and the flavon masses 
around the lower bound 
$v_T = 2m_{\varphi_{T1}} = m_{\varphi_{T2}} = m_{\varphi_{T3}} = 65$ GeV.
To calculate these values we use MadGraph5 \cite{MG5} with default 
momentum and rapidity cuts
\footnote{Because of the momentum and rapidity cuts, 
the values of production cross section are not the same as the values in Table 2 in the previous paper \cite{Muramatsu:2016bda}.}.
We find that the $\varphi_{T1}$ production cross sections are larger than other 
flavon production cross sections because $\varphi_{T1}$ is lighter than the other flavons.
Moreover $\varphi_{T1}$ production cross sections are $O(1)$ fb, and therefore 
we can expect the $A_4$ flavon signals at the LHC.

Because $\varphi_{T1}$ mainly decays into tau and anti-tau lepton, 
the expected flavon signal processes at proton - proton colliders include at least 
three tau leptons.
In collider experiments identification of tau leptons is more difficult than that of the  
other charged leptons.
Therefore we have to consider whether all tau leptons in flavon signal processes can be 
identified or not.
To do so we perform detector simulation by using MadGraph5 \cite{MG5}, 
Pythia \cite{Pythia}, and Delphes \cite{Delphes}.
We generate events for $pp\rightarrow \tau^- \tau^- \tau^+ \bar{\nu}_\tau$ and 
$\tau^- \tau^- \tau^+ \tau^+$ processes through $\varphi_{T1}$ production and 
perform detector simulation based on the LHC 
detector performance.
Then we count the number of events in which all tau leptons are identified as tau-tagged jets, and find that the expected number of event is less than one for both final-states when 3000 fb$^{-1}$ data are collected.
Therefore it is impossible to get sufficient discovery significance for our flavon at the LHC.
This feature is due to the low momentum cut for tau jets which come from the flavon decay\footnote{
Signals with leptonic decays of tau leptons are studied in
Ref.~\cite{Kanemura:2011kx}.
Because the charged leptons from tau decays tend to have a small
momentum fraction,
these signals are also suffered from the low momentum cut of leptons.}.
In low flavon mass region, although the flavon production cross section is large, 
charged leptons from the light flavon decay tend to have low momentum.

\section{Electron - photon collider}\label{sec:4}
\begin{figure}[htb]
\centering
\includegraphics[width=0.40\columnwidth]{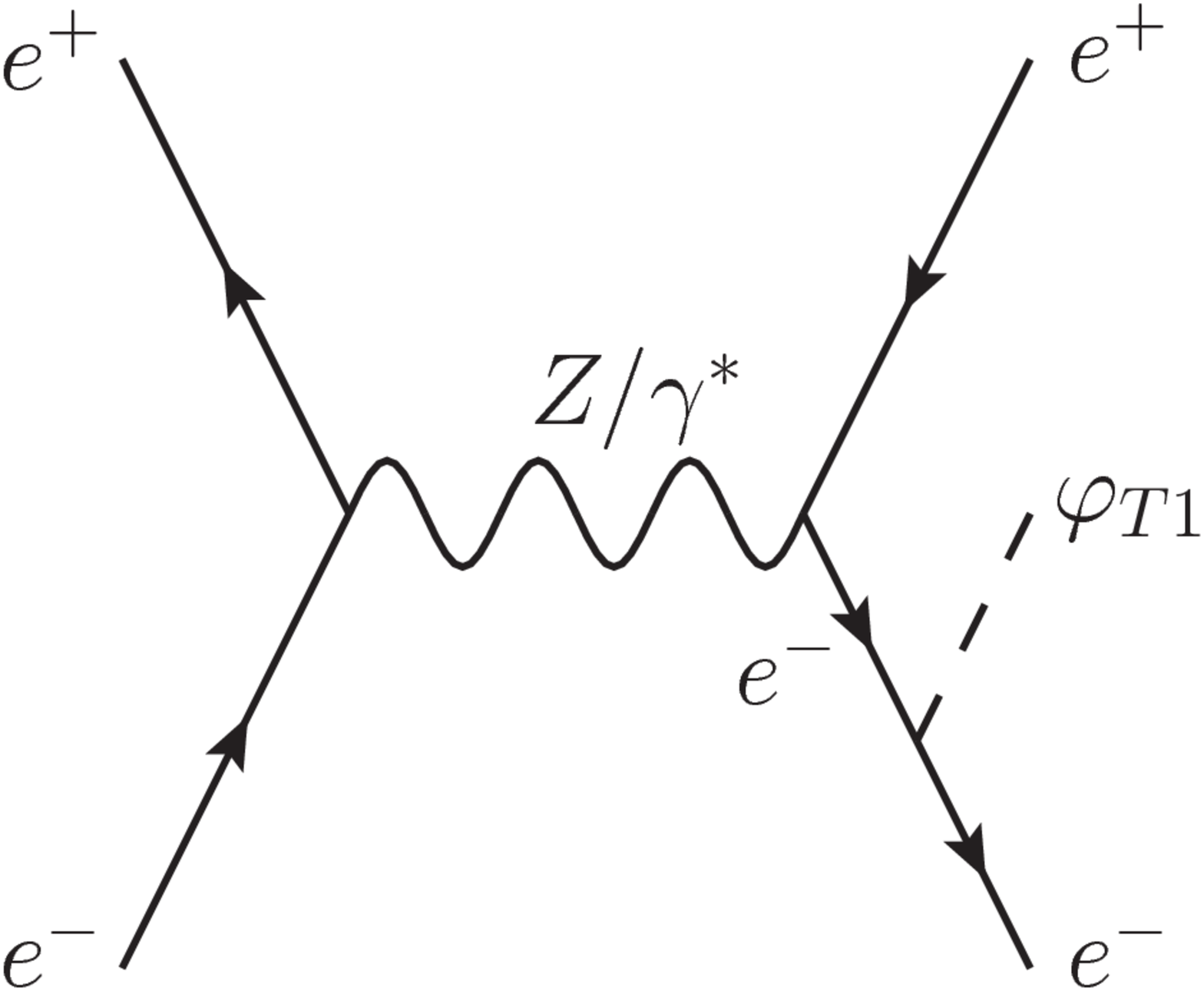}
\includegraphics[width=0.40\columnwidth]{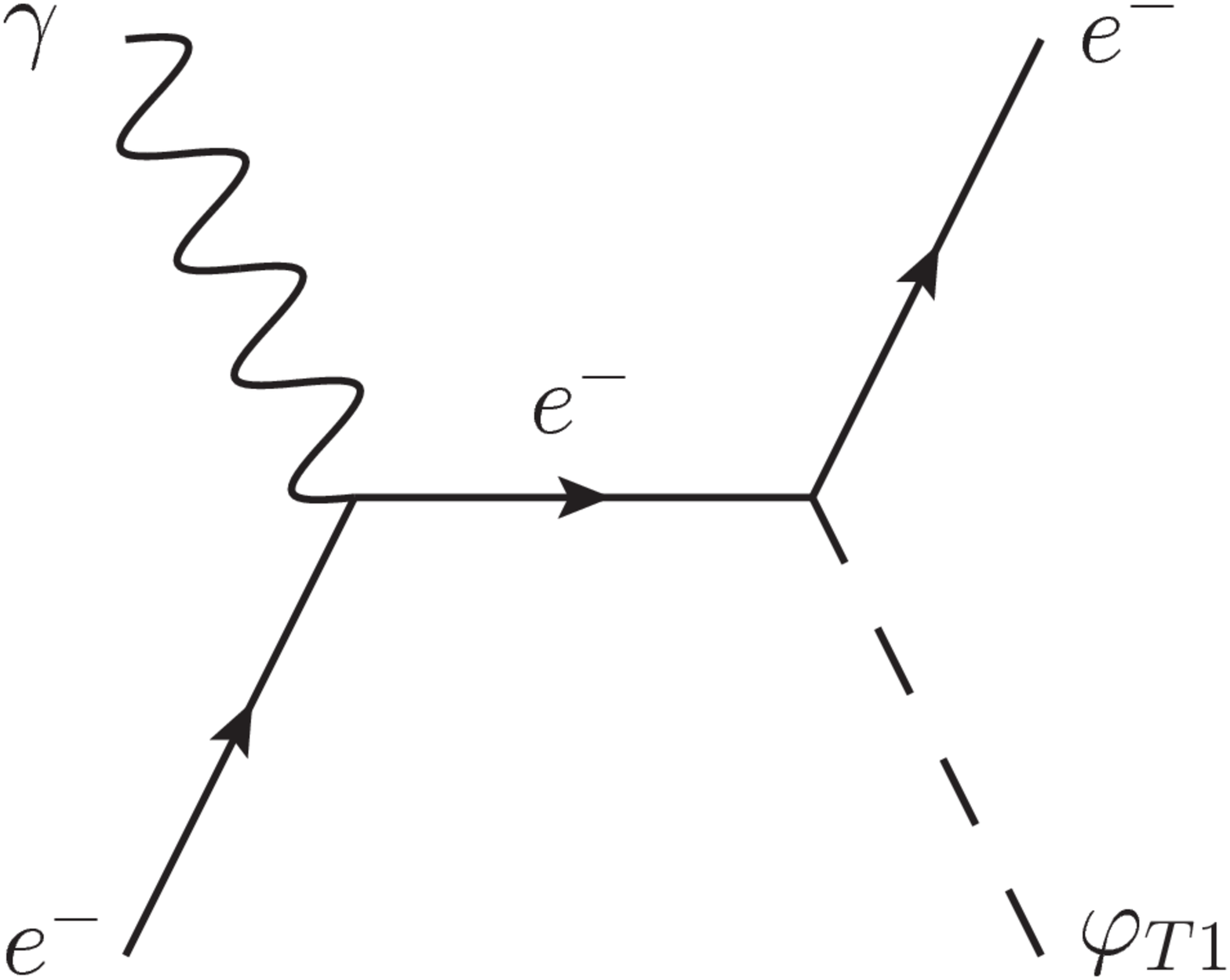}
\caption{
One of the diagram which produces the flavon $\varphi_{T1}$ 
at the electron - positron collider (left) and electron - photon collider (right).}
\label{fig:diagram}
\end{figure}

At lepton colliders, flavons are produced by $e^+ e^- \rightarrow l^+ l^- \varphi_T$ 
processes.
As with flavon production processes at the LHC, we find that 
it is hard to obtain sufficient discovery significance from these flavon processes at the ILC
\footnote{Another flavon signal at lepton colliders is a t-channel process, such as 
a $e^+ e^- \rightarrow l^+l^-$ process.
In the previous paper \cite{Muramatsu:2016bda} we discuss a constraint from this process at 
the LEP experiment \cite{LEPc} and showed this constraint is weaker than the constraint 
from $\tau$ lepton flavor violating decays.}.
As an alternative possibility for obtaining flavon signals at lepton
colliders, we consider an electron - photon collision as an option of
future lepton collider experiments such as the ILC.
At lepton colliders, a high-energy photon beam can be produced by
Compton backscattering of laser photons with electrons.
In electron - photon collisions, the lowest-order process for flavon
production is $e^-\gamma\to\ell^-\varphi_T$.
In Figure \ref{fig:diagram}, we show one of the diagrams for $\varphi_{T1}$ production at
the electron - positron collider (left) and at the electron
- photon collider (right).
Because of the order of QED interactions and the final-state phase-space volume,
we expect the flavon production cross section is larger at the
electron - photon collider than that at the electron - positron collider.
By including the decay of $\varphi_T\to l^+ l'^-$ the process contains
three leptons, all of which are expected to have larger momentum.

Here we discuss luminosity of electron - photon collision which could be realized at the ILC.  The photon beam can be obtained from an $e^-$ beam applying the energy spectrum of backscattered Compton photons~\cite{Ginzburg:1982yr, Ginzburg:1999wz}. The luminosity for $e^- e^-$ can be higher than the expected $e^+ e^-$ luminosity by a factor greater than 3~\cite{Telnov:1995hc, Fujii:2016raq}. Then the luminosity for the photon-photon collision is estimated to be $\mathcal{L}_{\gamma \gamma} \simeq \mathcal{L}_{e^- e^-} \times 3.6 \% $ where we assume a $3.6 \%$ decreasing effect from the photon energy distribution~\cite{Fujii:2016raq}-\cite{Richard:2016nhm}.  Since we have one photon beam for electron - photon collision we roughly guess the luminosity as $\mathcal{L}_{e^- \gamma} \sim \sqrt{0.036}  \mathcal{L}_{e^- e^-}  \sim 3 \sqrt{0.036} \mathcal{L}_{e^+ e^-}$. Therefore $\mathcal{L}_{e^- \gamma}$ would be around $50$-$60 \%$ of that of the electron-positron collision, and we simply apply the factor of $60 \%$ in our following analysis. For the collision energy, the energy distribution of the initial photon has a peak at around $80\%$ of the electron energy. In our analysis we simply assume the photon energy is $80 \%$ of the electron's one. Thus we consider electron - photon collision with beam energy 125$\times$100, 250$\times$200 and 500$\times$400 where the numbers before/after $\times$ indicate electron/photon energy in unit of [GeV]. 

\begin{table}[ht]
\begin{center}
\begin{tabular}{c||c|c|c|c|c|} 
& \multicolumn{2}{|c|}{ILC} & \multicolumn{3}{|c|}{upgraded ILC} \\ \hline
Beam energy [GeV$^2$] & 250$\times$200 & 125$\times$100 & 500$\times$400 &
250$\times$200 & 125$\times$100 \\
Luminosity [fb${}^{-1}$] & $300$ & $300$ & $4800$ &
2400 & 1200
\end{tabular}
\caption{The beam energies and luminosities to calculate flavon signals.}
\label{tab:BEIL}
\end{center}
\end{table}
In this paper we calculate flavon signals under the beam energy and luminosity which are 
sumarized in Table \ref{tab:BEIL}.
We estimate these numbers based on the planned beam energy and luminosity at the ILC 
\cite{ILC_scenarios}.

\subsection{Flavor conserving processes}\label{sec:5}
As the flavon Yukawa couplings which are proportional to $m_\tau$ induce large 
cross section, a process which we expect its observation is 
$e^- \gamma \rightarrow \tau^+ \tau^- e^-$.
This process mainly 
comes from $\varphi_{T3}$ flavon production 
$e^- \gamma \rightarrow \tau^- \varphi_{T3}$ followed by $\varphi_{T3}$ decays into 
$e^-$ and $\tau^+$.
SM background processes are  
$e^- \gamma \rightarrow e^- Z/\gamma^*$ followed by 
$Z/\gamma^* \rightarrow \tau^+ \tau^-$.
Therefore to reduce the background events, cuts on the invariant mass of a tau pair are useful.
Moreover the invariant mass of $e^-$ and $\tau^+$ shows significant evidence for 
this signal.

In Table \ref{tab:FCCS_BE} we summarize the cross section of 
the $e^- \gamma \rightarrow \tau^+ \tau^- e^-$ process for each beam energy.
In this calculation we use MadGraph5 \cite{MG5} and $v_T = 2m_{\varphi_{T1}} = m_{\varphi_{T2}} = m_{\varphi_{T3}} = 65$ GeV. 
In this calculation we use minimal $p_T=0.1$ GeV and maximal $|\eta|=4.0$ for the final-state charged leptons.
\begin{table}[ht]
\begin{center}
\begin{tabular}{c|ccc} \hline \hline
Beam energy [GeV$^2$] & 500$\times$400 & 250$\times$200 & 125$\times$100 \\
Signal cross section [fb] & $2.2$ & $8.6$ & $3.1 \times 10$ \\ 
Background cross section [fb] & $3.0 \times 10$ & $1.1 \times 10^2$ & 
$3.6 \times 10^2$ \\ \hline \hline 
\end{tabular}
\caption{The cross section of the $e^- \gamma \rightarrow 
\tau^+ \tau^- e^-$ process for each beam energy where 
$v_T = 2m_{\varphi_{T1}} = m_{\varphi_{T2}} = m_{\varphi_{T3}} = 65$ 
GeV.
In this calculation we use minimal $p_T=0.1$ GeV, and maximal $|\eta|=4.0$ for the final-state charged leptons.
}
\label{tab:FCCS_BE}
\end{center}
\end{table}
In Table \ref{tab:FCCS_BE}, this cross section becomes small as the beam energy becomes 
large, and thus the 500$\times$400 GeV$^2$ beam energy looks 
disadvantaged to detect this signal.

Hereinafter we study event selection cuts to obtain sufficient discovery significance.
We perform a detector simulation by using MadGraph5 \cite{MG5}, Pythia \cite{Pythia}, and Delphes \cite{Delphes}.
First we generate events for the signal process and SM background process,  
and perform the detector simulation based on the ILC detector proposal \cite{DSiD}
\footnote{In this paper we use a tau-tagging efficiency $40 \%$ as a default value
in Ref. \cite{DSiD}.
In reality, at the ILC the tau-tagging efficiency may be higher; for example in Ref. \cite{Baer:2013cma} the tau-tagging efficiency is quoted at $60 \%$.
The expected event numbers for the events with $n$ tau's would multiply by roughly $n$-th power of the enhancement rate of the tau-tagging efficiency.
}.
Next we count the number of events which satisfy our selection rules.
Our first selection rule is whether an event contain  two tau-jets $\tau_h$ and 
one electron $e^-$ or not.
Second selection rule is applying a cut on the invariant mass of the tau-jet pair $m_{\tau_h^- \tau_h^+}$ to reduce the events which come from the SM.
Finally to collect the events which come from the signal process, we use 
the invariant mass of $e^-$ and $\tau_h^+$ $m_{e^-\tau_h^+}$.

\begin{figure}[htb]
\centering
\includegraphics[width=0.49\columnwidth]{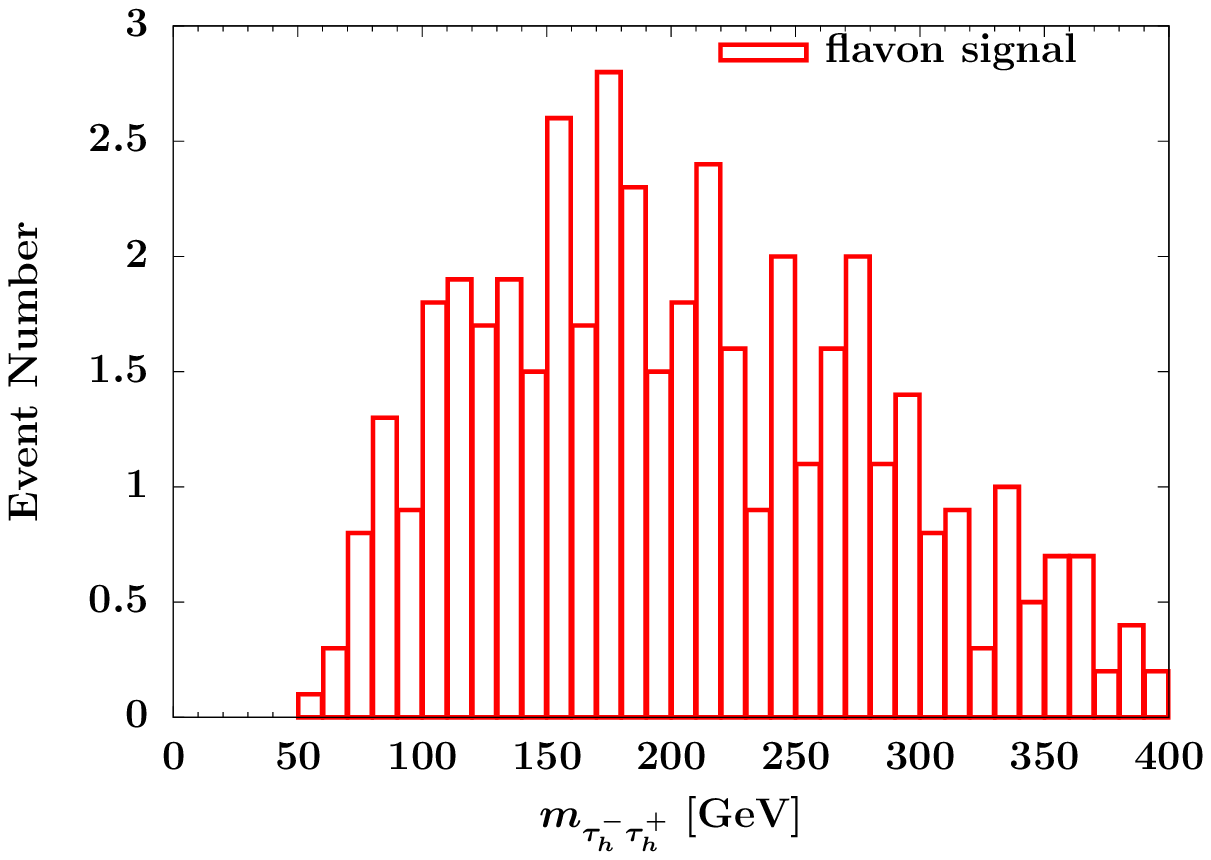}
\includegraphics[width=0.49\columnwidth]{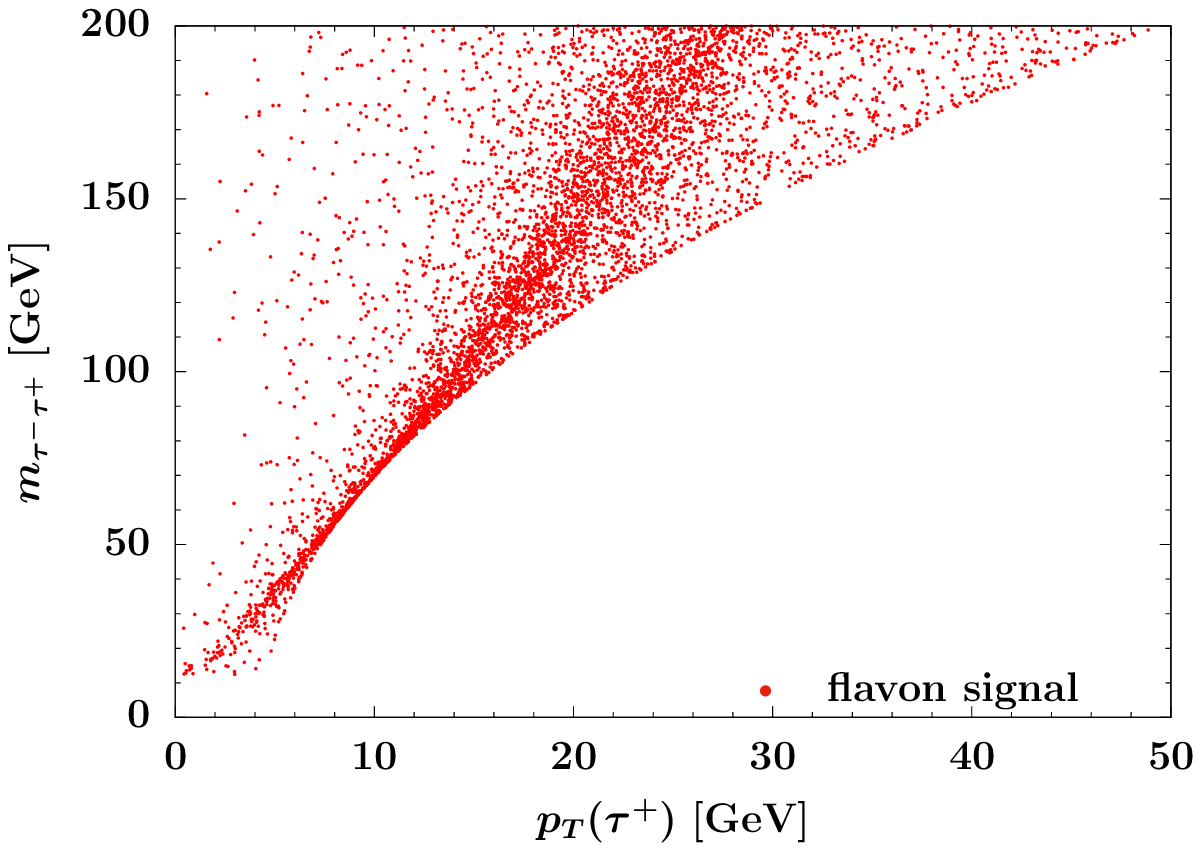}
\caption{
The $m_{\tau_h^-\tau_h^+}$ distribution after first selection (left) and 
the distribution where the horizontal axis is 
the transverse momentum for anti-tau lepton $p_T(\tau^+)$ and vertical axis is 
the $m_{\tau^-\tau^+}$ in a parton level simulation (right).
In these calculation we consider only signal process and generate 10 times expected number of events at the ILC.}
\label{fig:mtautau}
\end{figure}

The left panel of Figure \ref{fig:mtautau} shows the $m_{\tau_h^-\tau_h^+}$ distribution after the first selection where the beam energy is 250$\times$200 GeV$^2$.
In this calculation we consider only the signal process and generate 10 times the expected number of events at the ILC.
This distribution has a dump at 70 GeV.
This dump comes from the $p_T$ cut $p_T(\tau^+_h) \leq 10$ GeV.
The right panel of Figure \ref{fig:mtautau} shows the distribution where the horizontal axis is $p_T(\tau^+)$ and vertical axis is $m_{\tau^-\tau^+}$ in 
a parton level simulation.
As with the left plot we consider only the signal process and generate 10 times the expected number of events at the ILC.
This picture shows that $p_T(\tau^+)$ and $m_{\tau^-\tau^+}$ positively 
correlate because if $\tau^+$ travels in a same direction as $\tau^-$, $m_{\tau^-\tau^+}$ is small and $p_T(\tau^+)$ is 
small to realize the momentum conservation.
Therefore for $p_T(\tau^+)> 10$ GeV there are a few events which have small 
$m_{\tau^-\tau^+}$\footnote{Because this picture comes from the parton level simulation, 
$p_T(\tau^+)$ is a tau-lepton transverse momentum.
On the other hand in the detector level simulation transverse momentum cuts 
are applied to the tau-hadron.
Although there is a difference between the transverse momentum for the tau-lepton and 
tau-hadron, we can understand the effect of $p_T(\tau^+_h)$ cuts on 
$m_{\tau_h^-\tau_h^+}$.}.
On the other hand in the SM background process $m_{\tau_h^-\tau_h^+}$ distribution has a peak at around the Z boson mass $M_Z$.
Because of these reasons, our second selection rule works strongly to select signal events in our model.

In Table \ref{tab:FCcut} we summarize 
ratios of the signal event rate and  background event rate.
Here the event rate means the ratio of the event number after the cuts and event number before the cuts.
Moreover we summarize ratios of the signal event number and background event number, and 
significances of the signal after applying our selection cuts  
at the ILC and upgraded ILC.
As we calculate in Table \ref{tab:FCCS_BE} the flavon masses 
are fixed to $v_T = 2m_{\varphi_{T1}} = m_{\varphi_{T2}} = m_{\varphi_{T3}} = 65$ 
GeV in this calculation.
The background event number is suppressed strongly, and therefore 
we use a significance which is based on the Poisson distribution $S_{cL}$ \cite{Ball:2007zza}.
$S_{cL}$ is defined by 
\begin{equation}
S_{cl} = \sqrt{2 ((N_S + N_{BG}) \ln{(1 + N_S/N_{BG})} -N_S)},
\end{equation}
where $N_S$ is the signal event number, and $N_{BG}$ is the background event number.
\begin{table}[ht]
\begin{center}
\begin{tabular}{c|ccc} \hline \hline
250$\times$200 [GeV$^2$] & $R_S$ [\%]/$R_{BG}$ [\%] & $N_S$/$N_{BG}$ & $S_{cL}$ \\ \hline
contain $\tau^+_h \tau^-_h e^-$ & 1.7/1.8 & 45/590 & 1.8 \\ 
$m_{\tau_h^-\tau_h^+} > M_Z + 15.0 $ [GeV] & 1.6/0.15 & 41/49 & 5.3 \\ 
$ 30\,\text{[GeV]} < m_{e^-\tau_h^+} < 65\,\text{[GeV]}$ & 1.5/0.00093  & 
38/3.1 & 12 \\ \hline \hline 
125$\times$100 [GeV$^2$] & $R_S$ [\%]/$R_{BG}$ [\%] & $N_S$/$N_{BG}$ & $S_{cL}$ \\ \hline
contain $\tau^+_h \tau^-_h e^-$ & 1.8/1.9 & 170/2100 & 3.7 \\ 
$m_{\tau_h^-\tau_h^+} > M_Z + 5.0$ [GeV] & 1.0/0.089 & 96/96 & 8.6 \\ 
$ 25\,\text{[GeV]} < m_{e^-\tau_h^+} < 70\,\text{[GeV]}$ & 1.0/0.028 & 
95/31 & 13 \\ \hline \hline 
500$\times$400 [GeV$^2$] & $R_S$ [\%]/$R_{BG}$ [\%] & $N_S$/$N_{BG}$ & $S_{cL}$ \\ \hline
contain $\tau^+_h \tau^-_h e^-$ & 0.91/1.4 & 97/2100 & 2.1 \\ 
$m_{\tau_h^-\tau_h^+} > 290$ [GeV] & 0.68/0.12 & 72/170 & 5.2 \\
$ 25\,\text{[GeV]} < m_{e^-\tau_h^+} < 70\,\text{[GeV]}$ & 0.67/0.0036 & 
72/5.2 & 17 \\ \hline \hline 
\end{tabular}
\caption{Ratios of the signal event rate and background event rate ($R_S$/$R_{BG}$),  
ratios of the signal event number and background event number ($N_S$/$N_{BG}$), and 
significances of the signal $S_{cL}$ after applying our selection rules 
at the ILC and upgraded ILC.
In this calculation we use 
$v_T = 2m_{\varphi_{T1}} = m_{\varphi_{T2}} = m_{\varphi_{T3}} = 65$ GeV.
}
\label{tab:FCcut}
\end{center}
\end{table}
In our second selection rule we remove events with larger $m_{\tau_h^-\tau_h^+}$  as the beam energy becomes larger.
This is because the $m_{\tau_h^-\tau_h^+}$ dump position which we showed in Figure \ref{fig:mtautau} becomes higher as 
the beam energy becomes larger.
Table \ref{tab:FCcut} shows that when the beam energy is 250$\times$200 GeV$^2$ and 
125$\times$100 GeV$^2$ we can expect sufficient discovery significance for the flavon mass around the lower limit.
When the beam energy is 500$\times$400 GeV$^2$, the discovery significance 
is sufficient even though the signal cross section is small, because 
at the upgraded ILC we can expect large integrated luminosity.

\begin{figure}[hbt]
\centering
\includegraphics[width=0.49\columnwidth]{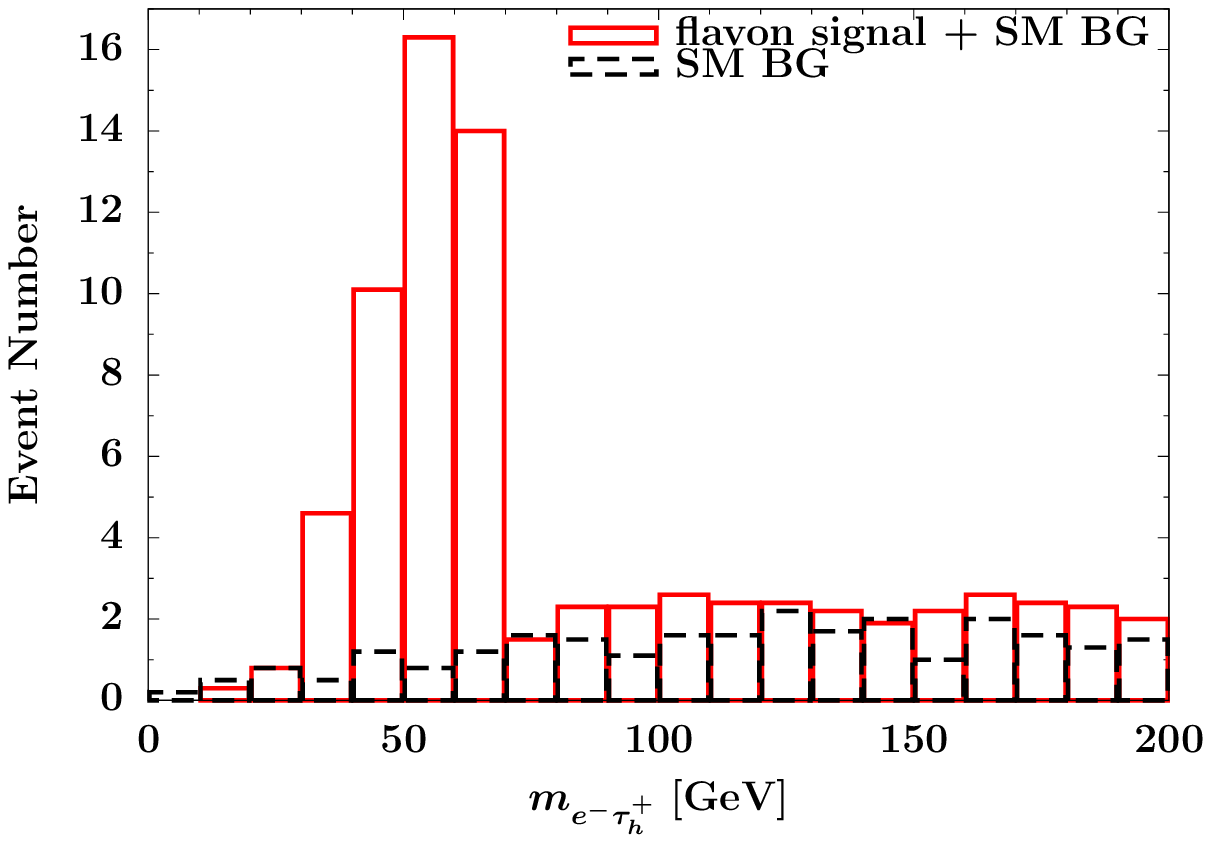}
\includegraphics[width=0.49\columnwidth]{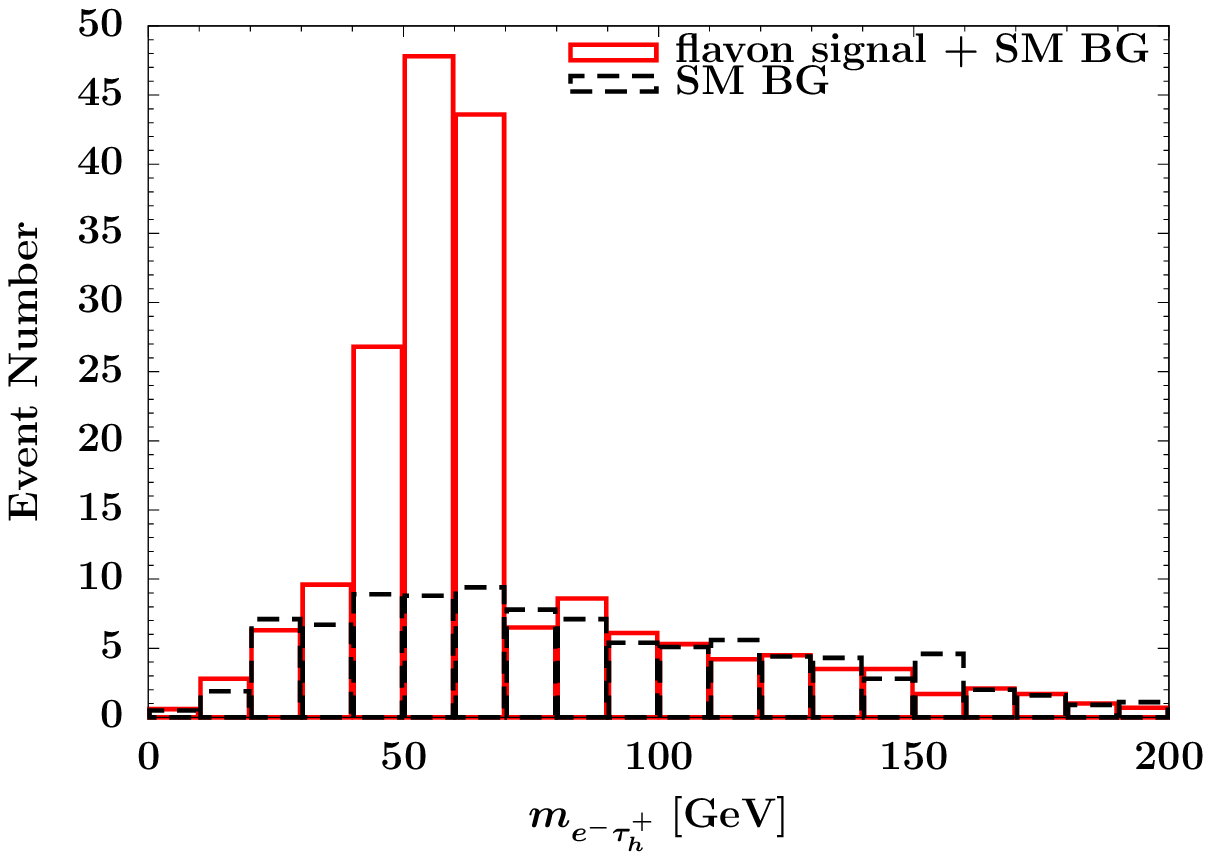}
\caption{
The $m_{e^-\tau_h^+}$ distribution  after the second selection rule
for the beam energy 250$\times$200 GeV$^2$ (left) and 125$\times$100 GeV$^2$ (right) 
at the ILC.
The red solid histograms are for both the SM and flavon processes, while the black dashed 
histograms are for only the SM process.
In this calculation we fix the flavon masses to be 
$v_T = 2m_{\varphi_{T1}} = m_{\varphi_{T2}} = m_{\varphi_{T3}} = 65$ GeV.
}
\label{fig:meletau1}
\end{figure}

\begin{figure}[bth]
\centering
\includegraphics[width=0.49\columnwidth]{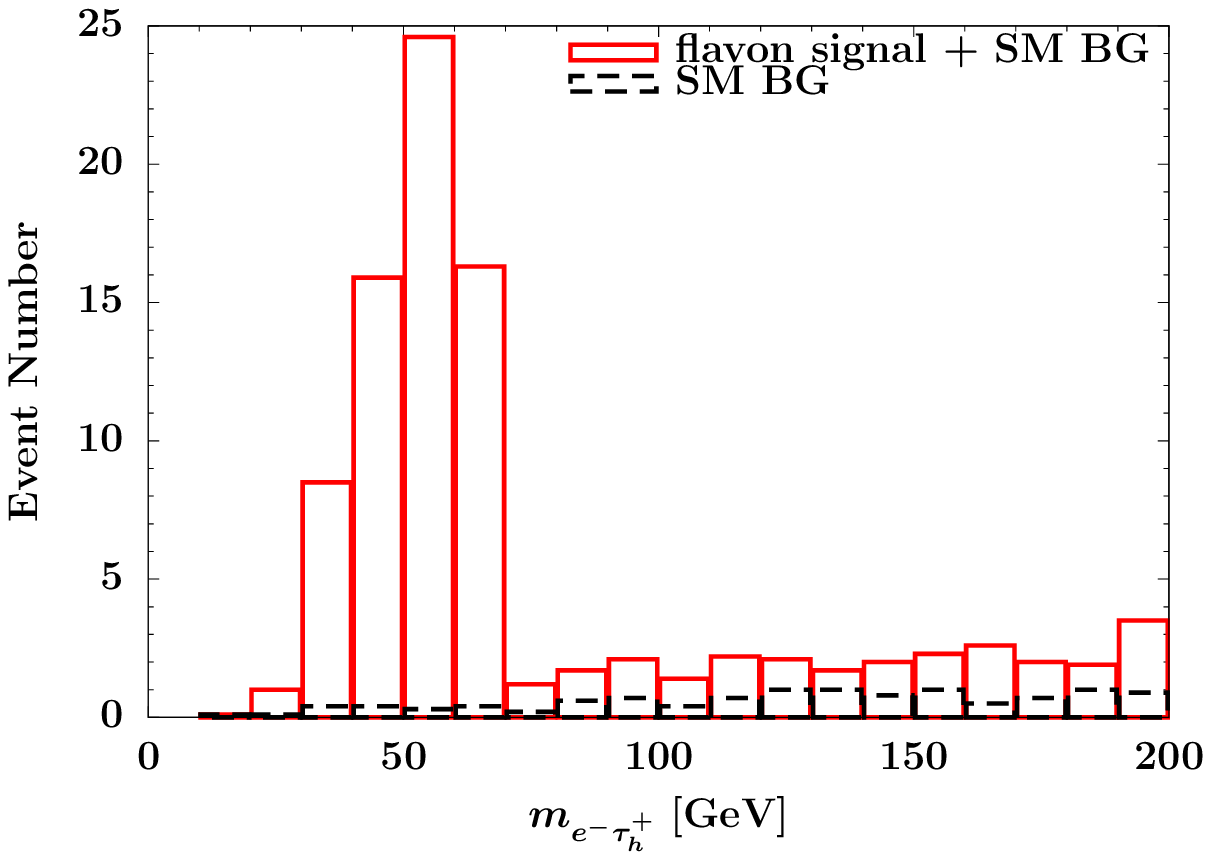}
\caption{
The $m_{e^-\tau_h^+}$ distribution where the beam energy is 
500$\times$400 GeV$^2$ at the upgraded ILC.
The red solid histograms indicate the number of events which come from the SM and flavon interactions, 
and the black dashed histograms indicate the number of events which come from 
the only SM interactions.
In this calculation the flavon masses are 
$v_T = 2m_{\varphi_{T1}} = m_{\varphi_{T2}} = m_{\varphi_{T3}} = 65$ GeV.}
\label{fig:meletau2}
\end{figure}

The left and right panels in Figure \ref{fig:meletau1} 
show the $m_{e^-\tau_h^+}$ distribution  after the second selection rule
for the beam energy 250$\times$200 GeV$^2$ and 125$\times$100 GeV$^2$ at the ILC, respectively.
The red solid histograms are for both the SM and flavon processes, while the black dashed 
histograms are for only the SM process.
In this calculation we fix the flavon masses to be 
$v_T = 2m_{\varphi_{T1}} = m_{\varphi_{T2}} = m_{\varphi_{T3}} = 65$ GeV.
Both figures show a clear peak around the flavon mass.
Figure \ref{fig:meletau2} is the same as Figure \ref{fig:meletau1}, but for the beam energy 
of 500$\times$400 GeV$^2$ at the upgraded ILC.
This figure also show a clear peak around the flavon mass.
From Table \ref{tab:FCcut} the significance for the beam energy 500$\times$400 GeV$^2$ 
is not larger than those for the beam energy 250$\times$200 GeV$^2$ and 125$\times$100 GeV$^2$ 
after second selection rule.
On the other hand the peak around the flavon mass for the beam energy 500$\times$400 GeV$^2$ is clearer than those for the beam energy 250$\times$200 GeV$^2$ and 125$\times$100 GeV$^2$.
This is because the number of events in small $m_{e^-\tau^+}$ region is tiny in the SM for the beam energy 500$\times$400 GeV$^2$.
This feature comes from the anti-tau-hadron transverse momentum cuts 
as we showed above.

\begin{figure}[hbt]
\centering
\includegraphics[width=0.49\columnwidth]{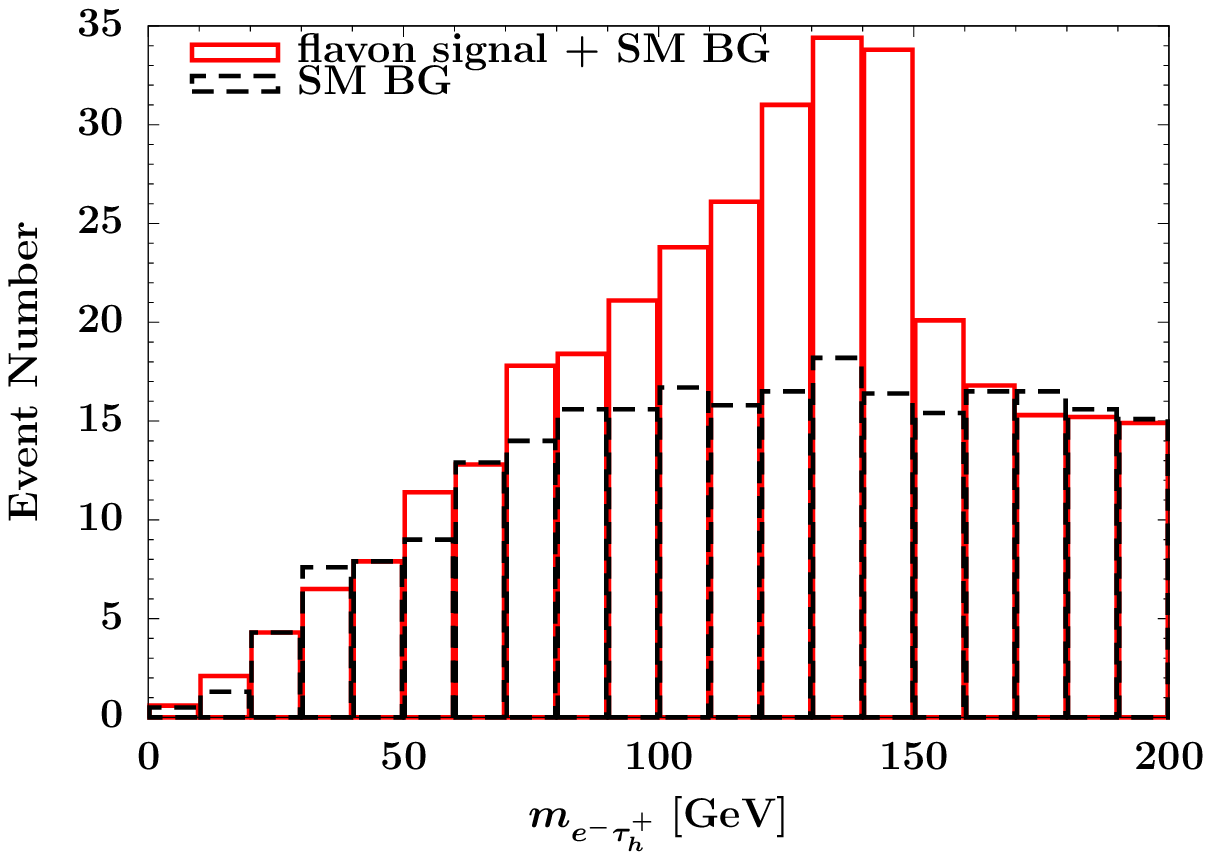}
\includegraphics[width=0.49\columnwidth]{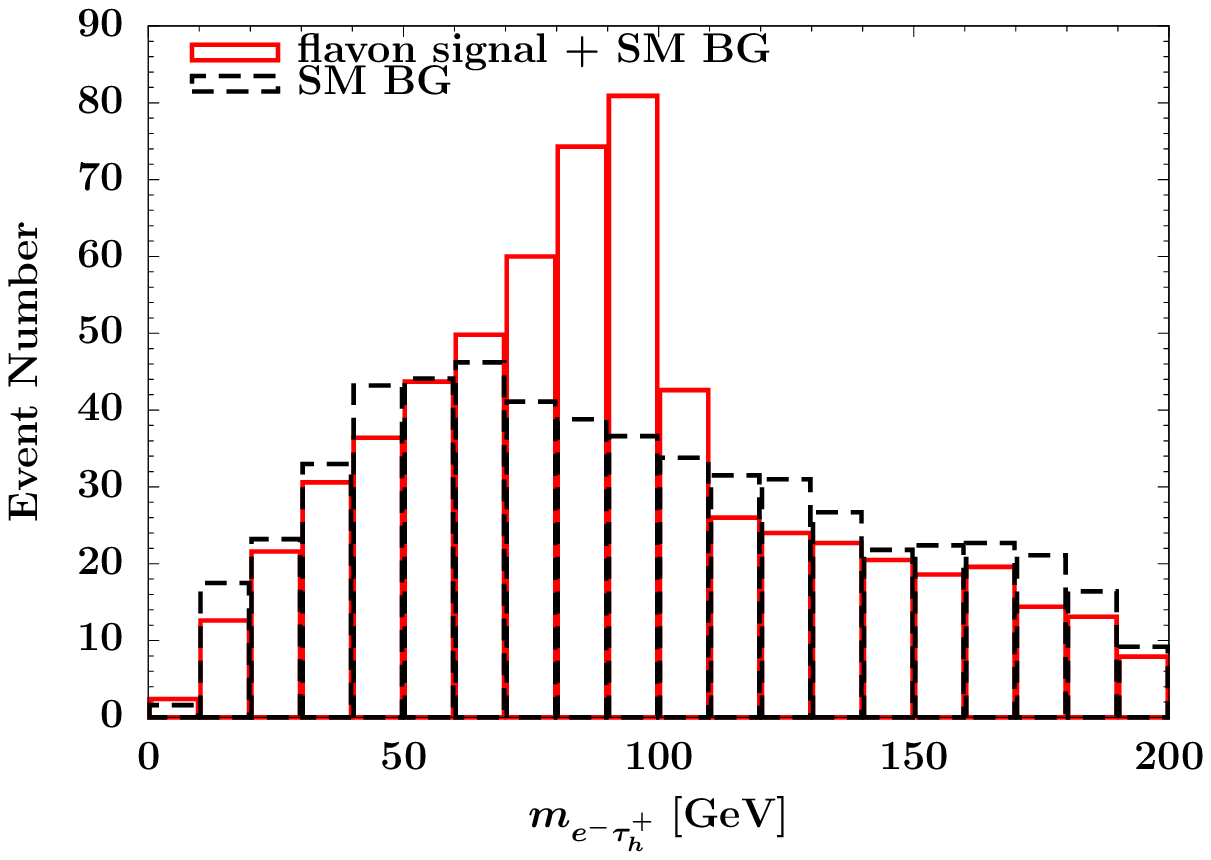}
\caption{
The $m_{e^-\tau_h^+}$ distribution where the beam energy is 
250$\times$200 GeV$^2$ (left) and 125$\times$100 GeV$^2$ (right) at the upgraded ILC.
The red solid histograms indicate the number of events which come from the SM and flavon interactions, 
and the black dashed histograms indicate the number of events which come from 
the only SM interactions.
In this calculation the flavon masses are 
$v_T = 2m_{\varphi_{T1}} = m_{\varphi_{T2}} = m_{\varphi_{T3}} = 150$ GeV (left) and 
100 GeV (right).
}
\label{fig:meletau3}
\end{figure}

Finally we study the discovery significance for different flavon masses
at the luminosity upgraded ILC case.
We consider the flavon masses 
$v_T = 2m_{\varphi_{T1}} = m_{\varphi_{T2}} = m_{\varphi_{T3}} = 100,\,150,\,
\text{and}\,200$ GeV.
When the beam energy is 250$\times$200 GeV$^2$, we find sufficient discovery significance for 
the flavon masses 100 GeV and 150 GeV.
When the beam energy is 125$\times$100 GeV$^2$, we find sufficient discovery significance for 
the flavon masses 100 GeV.
The Figure \ref{fig:meletau3} shows the $m_{e^-\tau_h^+}$ distribution 
at the luminosity upgraded ILC.
The left panel is for the flavon masses 150 GeV and the beam energy 250$\times$200 GeV$^2$, 
and the right panel is for the flavon masses 100 GeV and the beam energy 125$\times$100 GeV$^2$.
We expect sufficient discovery significance for the flavon masses lighter than 150 GeV.

\clearpage

\subsection{Flavor violating processes}\label{sec:6}
In our model, there are following three lepton flavor violating processes: 
$e^- \gamma \rightarrow \tau^+ \mu^- \mu^-$, $\mu^+ \tau^- \tau^-,$ and $e^+ \tau^- \mu^-$.
Amplitudes for these flavor violating processes are proportional to 
at least two charged lepton masses.
The cross section for the $e^+ \tau^- \mu^-$ final-state is smaller than 
the cross sections for the other two final-states, because 
this process comes from the coupling which is proportional to $m_e$.
Therefore in this paper we consider 
$e^- \gamma \rightarrow \tau^+ \mu^- \mu^-$ and 
$e^- \gamma \rightarrow \mu^+ \tau^- \tau^-$ processes.
The cross sections for both final-states are nearly same.
Because these processes induce the lepton flavor violation, 
SM contributions are extremely suppressed.
Therefore even if we detect a few events for these final-states, 
it can be a strong evidence for our model. 

In Table \ref{tab:FVCS_BE} we summarize the cross section of 
the $e^- \gamma \rightarrow \tau^+ \mu^- \mu^-$ process for 
each beam energy.
Here we adopt $v_T = 2m_{\varphi_{T1}} = m_{\varphi_{T2}} = m_{\varphi_{T3}}
= 65$ GeV.
The cross sections are calculated by using MadGraph5 \cite{MG5} with kinematical cuts: 
minimal $p_T$ is 0.1 GeV, and maximal $|\eta|$ is 4.0 for the final-state charged leptons.
\begin{table}[ht]
\begin{center}
\begin{tabular}{c|ccc} \hline \hline
Beam energy [GeV$^2$] & 500$\times$400 & 250$\times$200 & 125$\times$100 \\
Cross section [fb] & $7.9 \times 10^{-3}$ & $3.1 \times 10^{-2}$ & 
$1.1 \times 10^{-1}$ \\ \hline \hline 
\end{tabular}
\caption{The cross section of the $e^- \gamma \rightarrow 
\tau^+ \mu^- \mu^-$ process for each beam energy where 
$v_T = 2m_{\varphi_{T1}} = m_{\varphi_{T2}} = m_{\varphi_{T3}} = 65$ 
GeV.
In this calculation we use minimal $p_T=0.1$ GeV, and maximal $|\eta|=4.0$ for the final-state charged leptons.}
\label{tab:FVCS_BE}
\end{center}
\end{table}
Table \ref{tab:FVCS_BE} shows that around the flavon lower mass limit 
the 250$\times$200 GeV$^2$ and 125$\times$100 GeV$^2$ beam energies are 
favored for detecting the flavor violating processes.
Moreover when $O(10-100)$ fb$^{-1}$ data are collected, 
we will detect these lepton flavor violating processes at the ILC.

As we mentioned in Section \ref{sec:3}, the identification of tau leptons is not straightforward.
Our signal processes have tau lepton, and therefore we perform detector simulation.
We generate $10^4$ events for each process by MadGraph5 \cite{MG5} 
and select events with three charged leptons and lepton flavor violation.
For the detector simulation we use Pythia \cite{Pythia} and Delphes \cite{Delphes} based on the ILC detector proposal \cite{DSiD}.
In Table \ref{tab:FVES} we summarize event rates and expected event numbers 
for each process at the ILC and upgraded ILC.
In this table we consider the 250$\times$200 GeV$^2$ and 125$\times$100 GeV$^2$ 
beam energies and $v_T =  
2m_{\varphi_{T1}} = m_{\varphi_{T2}} = m_{\varphi_{T3}} = 65$ GeV.
In the $\mu^+ \tau^- \tau^-$ mode there are final-states which 
include $e^-$ such as $\mu^+ \tau^- e^-$ and $\mu^+ e^- e^-$.
However these final-states have large SM background which 
comes from flavor conserving processes $e^- \gamma \rightarrow \tau^+ \tau^- e^-$.
Therefore in this calculation we ignore these final-states.
\begin{table}[tbh]
\begin{center}
\begin{tabular}{c|c||c|ccc}
 & \shortstack{Cross \\ section [fb]} & 
\shortstack{Final \\ state}
 & \shortstack{Event \\ rate [\%]} & 
\shortstack{Event \\ number at \\ 300(300) [fb$^{-1}$]} & 
\shortstack{Event \\ number at \\ 2400(1200) [fb$^{-1}$]} \\ \hline
$\tau^+ \mu^- \mu^-$ & $3.1 \times 10^{-2}$ & sum & 
19(23) & 1.8(7.4) & 14(30) \\
mode & $(1.1 \times 10^{-1})$ & $\tau_h^+ \mu^- \mu^-$ & 
6.5(8.9) & 0.59(2.9) & 4.8(12) \\
 & & $\mu^+ \mu^- \mu^-$ & 6.5(7.2) & 0.59(2.3) & 4.7(9.3) \\
 & & $e^+ \mu^- \mu^-$ & 6.5(6.8) & 0.60(2.2) & 4.8(8.8) \\ \hline \hline
$\mu^+ \tau^- \tau^-$ & $3.1 \times 10^{-2}$ & sum & 
6.5(7.1) & 0.59(2.3) & 4.8(9.2) \\
mode & $(1.1 \times 10^{-1})$ & $\mu^+ \tau_h^- \mu^-$ & 
3.2(3.5) & 0.29(1.1) & 2.3(4.6) \\
 & & $\mu^+ \tau_h^- \tau_h^-$ & 2.2(2.3) & 0.20(0.76) & 1.6(3.0) \\
 & & $\mu^+ \mu^- \mu^-$ & 1.1(1.2) & 0.10(0.40) & 0.84(1.6) \\
\end{tabular}
\caption{Event rates and expected event numbers for each final-state where 
beam energy is 250$\times$200 GeV$^2$ (125$\times$ 100 GeV$^2$) and $v_T =  
2m_{\varphi_{T1}} = m_{\varphi_{T2}} = m_{\varphi_{T3}} = 65$ GeV
at the ILC and luminosity upgraded ILC.}
\label{tab:FVES}
\end{center}
\end{table}
This table shows that around the flavon mass lower limit observations of the lepton flavor violating signals are expected at the ILC.
Moreover at the upgraded ILC we can expect the observations of 
about 20 (40) flavor violating events for 250$\times$200 GeV$^2$ (125$\times$ 100 GeV$^2$) beam energy.
If many flavor violating events are observed, we can test this model by using a ratio of the event number for each process.

In this subsection we do not consider background events from the SM interactions.
Because these signals have 
large lepton flavor violation, for example the $e^- \gamma \rightarrow \tau^+ \mu^- \mu^-$ process induces $\Delta N_e = -1$, $\Delta N_\mu = +2$, and $\Delta N_\tau = -1$.
In the SM lepton flavor violations are induced via the weak interactions, and 
the weak interactions produce neutrinos.
Therefore even if the number of background events from the SM interactions is sizable, we can identify signals by using a missing momentum cut.

\begin{figure}[bth]
\centering
\includegraphics[width=0.7\columnwidth]{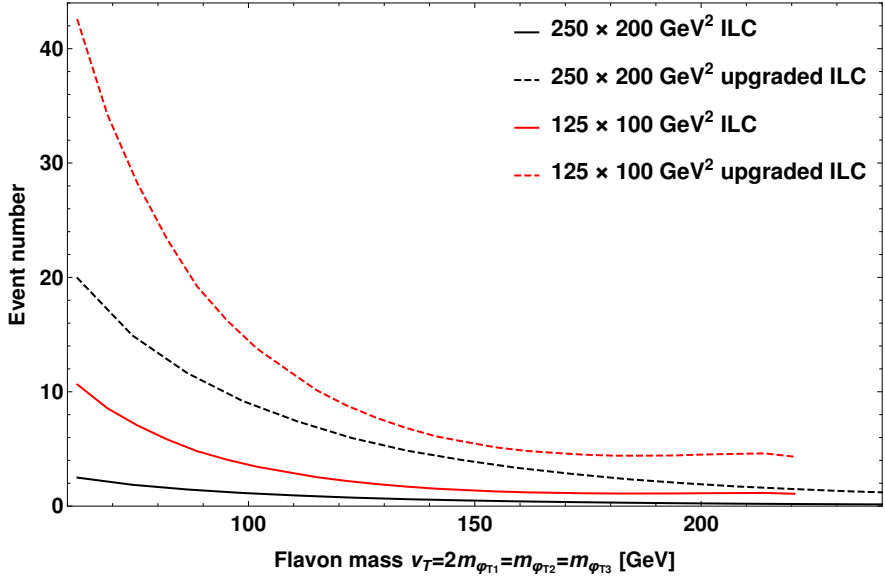}
\caption{
The flavon mass dependence for the event number of
the lepton flavor violating process where the beam energy and luminosity is 
250$\times$200 GeV$^2$ at the ILC (black solid line), 
125$\times$100 GeV$^2$ at the ILC (red solid line), 
250$\times$200 GeV$^2$ at the upgraded ILC (black dashed line), and 
125$\times$100 GeV$^2$ at the upgraded ILC (red dashed line).}
\label{fig:EN3LFV}
\end{figure}

Figure \ref{fig:EN3LFV} shows flavon mass dependence for the sum of the event number for the flavor violating signals.
In this calculation we assume beam energy is 250$\times$200 GeV$^2$ and 
125$\times$100 GeV$^2$.
The black solid line, red solid line, black dashed line, and 
red dashed line shows the event number for the 250$\times$200 GeV$^2$ beam energy at the ILC, 
125$\times$100 GeV$^2$ beam energy at the ILC, 
250$\times$200 GeV$^2$ beam energy at the upgraded ILC, and 
125$\times$200 GeV$^2$ beam energy at the upgraded ILC, respectively. 
The red lines have a dump at around $m_{\varphi_{T2}} = m_{\varphi_{T3}} = 225$ GeV, 
because of the loss of the phase space.
We can expect more event number for the 125$\times$ 100 GeV$^2$ beam energy than 
that for the 250$\times$ 200 GeV$^2$ beam energy, while we can search for wider 
mass regions with the 250$\times$ 200 GeV$^2$ beam energy.

If we assume the number of background events is 1, the minimal number of signal events which gives $S_{cL}=3$ is 4.
We estimate mass upper limits which satisfy $S_{cL}=3$ for any flavon mass.
At the ILC case it is not possible to 
satisfy $S_{cL}=3$ when the beam energy is 250$\times$200 GeV$^2$.
On the other hand when 
the beam energy is 125$\times$100 GeV$^2$ it is possible to satisfy $S_{cL}=3$ below 
the flavon mass $m_{\varphi_{T2}} = m_{\varphi_{T3}}=96$ GeV.
In the upgraded ILC case when the beam energy is 250$\times$200 GeV$^2$, it is possible to 
satisfy $S_{cL}=3$ below the flavon mass $m_{\varphi_{T2}} = m_{\varphi_{T3}}=150$ GeV.
Moreover when 
the beam energy is 125$\times$100 GeV$^2$ it is possible to satisfy $S_{cL}=3$ below the 
cross section dump at around flavon mass $m_{\varphi_{T2}} = m_{\varphi_{T3}}=225$ GeV.

\section{Summary and discussion}\label{sec:7}
In this paper, we discussed collider signals of the light flavons $\varphi_T$ which are introduced in the modified AF model.
At the LHC and in the electron - positron collision at the ILC, we can not obtain sufficient discovery significance.
We found that the electron - photon collision at the ILC 
has advantages to search for the signals where the flavons are 
produced as $e^-\gamma \to l^- \varphi_T$ and decay into 
two charged leptons.
Then we analyzed signals of the flavor-conserving final-state 
$\tau^+ \tau^- e^-$, and flavor-violating final-states 
$\tau^+ \mu^- \mu^-$ and $\mu^+ \tau^- \tau^-$ by carrying out 
numerical simulation.
For the former final-state, SM background can be strongly suppressed by 
imposing cuts on the invariant masses of the final-state leptons.
As a result around the flavon mass lower limit we can obtain sufficient discovery significance for each beam energy.
Moreover at the upgraded ILC we expect sufficient discovery significance for the flavon masses lighter than 150 GeV.
For the later final-states, we found that at the ILC we can expect signal observation,  and at the upgraded ILC we expect sufficient signal observation to test this model.
In particular at the upgraded ILC we expect sufficient discovery significance for the flavon masses lighter than 225 GeV.
Therefore stronger flavon mass constraints are obtained from the future collider experiments compared to those from the flavor violating decay of leptons.

\clearpage

\vspace{0.5cm}
\noindent
{\bf Acknowledgement}

YM is supported in part by the National Natural Science Foundation of China (NNSFC) under contract Nos. 11435003, 11225523, and 11521064.
YM and YS are supported in the part by the National Research Foundation of Korea (NRF) Research Grant NRF - 2015R1A2A1A05001869.
YS is supported in part by JSPS Grant-in-Aid for Scientific Research No. 16J05332.

\end{document}